\documentclass[conference]{IEEEtran}
\IEEEoverridecommandlockouts
\usepackage{algorithm}
\usepackage{algpseudocode}
\usepackage{amsfonts}
\usepackage{bm}
\usepackage{amsmath}
\usepackage{cite}
\usepackage{float}
\usepackage{amssymb}
\usepackage{enumerate}
\usepackage{color,xcolor}
\usepackage{graphicx}
\usepackage{multirow,tabularx}
\usepackage{subfigure}
\usepackage{hyperref}

\makeatletter

\newcommand{\Rmnum}[1]{\expandafter\@slowromancap\romannumeral #1@}
\makeatother

\usepackage{bookmark}

\ifCLASSINFOpdf

\else

\fi

\begin{document}

\title{\huge{Sparse Vector Recovery: Bernoulli-Gaussian Message Passing}} %Bernoulli-Gaussian, Hybrid,
\author{\IEEEauthorblockN{Lei Liu\IEEEauthorrefmark{1}\IEEEauthorrefmark{3}, Chongwen Huang\IEEEauthorrefmark{1}, Yuhao Chi\IEEEauthorrefmark{2}\IEEEauthorrefmark{3}, Chau Yuen\IEEEauthorrefmark{1}, Yong Liang Guan\IEEEauthorrefmark{2}, and Ying Li\IEEEauthorrefmark{3}\\
\IEEEauthorrefmark{1}Singapore University of Technology and Design, Singapore\\
\IEEEauthorrefmark{2}Nanyang Technological University, Singapore}
\IEEEauthorrefmark{3}State Key Lab of ISN, Xidian University, China
%\\E-mail:yli@mail.xidian.edu.cn}

%\author{\IEEEauthorblockN{\normalsize Lei Liu, Chongwen Huang, Yuhao Chi, Chau Yuen, Yong Liang Guan and Ying Li}

%\author{\IEEEauthorblockN{Lei Liu, Chongwen Huang, Chau Yuen, Yong Liang Guan and Ying Li}
%
%

\thanks{This work was supported in part by the National Natural Science Foundation of China under Grants 61671345, and in part by the Singapore A*STAR SERC Project under Grant 142 02 00043. The first author was also supported by the China Scholarship Council under Grant 20140690045. }
}

\maketitle

\begin{abstract}
Low-cost \emph{message passing} (MP) algorithm has been recognized as a promising technique for sparse vector recovery. However, the existing MP algorithms either focus on \emph{mean square error} (MSE) of the value recovery while ignoring the sparsity requirement, or \emph{support error rate} (SER) of the sparse support (non-zero position) recovery while ignoring its value. A novel low-complexity \emph{Bernoulli-Gaussian MP} (BGMP) is proposed to perform the value recovery as well as the support recovery. Particularly, in the proposed BGMP, support-related Bernoulli messages and value-related Gaussian messages are jointly processed and assist each other. In addition, a strict lower bound is developed for the MSE of BGMP via the \emph{genie-aided minimum mean-square-error} (GA-MMSE) method. The GA-MMSE lower bound is shown to be tight in high signal-to-noise ratio. Numerical results are provided to verify the advantage of BGMP in terms of final MSE, SER and convergence speed.
%Low-cost message passing (MP) algorithm has been recognized as a promising technique for sparse vector recovery. However, the exsiting MP algorithms either focus on mean square error (MSE) of the value recovery while ignoring the sparsity requirement, or support error rate (SER) of the sparse support (non-zero position) recovery while ignoring its value. A novel low-complexity Bernoulli-Gaussian MP (BGMP) is proposed to perform the value recovery as well as the support recovery. Particularly, in the proposed BGMP, support-related Bernoulli messages and value-related Gaussian messages are jointly processed and assist each other. In addition, a strict lower bound is developed for the MSE of BGMP via the genie-aided minimum mean square error (GA-MMSE) method. The GA-MMSE lower bound is shown to be tight in high signal noise ratio. Numerical results are provided to verify the advantage of BGMP in terms of final MSE, SER and convergence speed.
\end{abstract}

\begin{IEEEkeywords}%\vspace{-0.3cm}
Bernoulli-Gaussian, belief propagation, compressed sensing, sparse vector recovery, factor graph.
\end{IEEEkeywords}

\IEEEpeerreviewmaketitle
%\newpage
\section{Introduction}
Recently, with the rapid development of the wireless network, we have entered the age of ``Big Data". Practically, most interesting data is typically sparse, and thus sparse vector recovery problems have attracted much interest in many engineering fields \cite{Eldar2012}, such as data collection, network monitoring, mmWave channel estimation, \emph{interest of things} (IoT), \emph{machine to machine} (M2M) communications, machine learning, \emph{cloud-radio access network} (C-RAN), etc.

Sparse vector recovery is a technique for reconstructing a sparse vector $\mathbf{x}=[x_1,\cdots,x_{K}]^T$ from an underdetermined noisy measurement $\mathbf{y}\in {\mathbb{R}}^{M \times 1}$:\vspace{-0.2cm}
\begin{equation}
\mathbf{y}=\mathbf{H}\mathbf{x}+\mathbf{n},\vspace{-0.2cm}
\end{equation}
where $\mathbf{H}\in {\mathbb{R}}^{M \times K}$ is a given measurement matrix, and $\mathbf{n}\sim\mathcal{N}^{M} (0,\sigma_n^2)$ a vector of independent \emph{additive white Gaussian noise} (AWGN). It is based on the principle that the sparsity is exploited to achieve a more efficient sampling than the classical \emph{Shannon-Nyquist} scheme \cite{Candes2006,Donoho2006}.

In the past decades, many sparse vector recovery algorithms have been proposed. One of the most popular schemes is formulated as the minimization of the squared error $\|\textbf{y}-\textbf{H}\hat{\textbf{x}}\|^2$ (where $\|\!\cdot\!\|$ denotes the Euclidean norm) under the constraint that the $l_0$ pseudo-norm of $\hat{\textbf{x}}$ is small. However, it is well known to be a NP-complete problem \cite{Natarajan1995}. Another well-known approach is LASSO \cite{Donoho2008}, where $l_0$-norm has been relaxed to the $l_1$-norm minimization problem:\vspace{-0.2cm}
\begin{equation}
\hat{\mathbf{x}}=\arg \mathop {\min }\limits_{\hat{\mathbf{x}}} \;\;\|\textbf{y}-\textbf{H}\hat{\textbf{x}}\|_2^2+\lambda\|\hat{\mathbf{x}}\|_1,\vspace{-0.2cm}
\end{equation}
which is convex and can be efficiently solved. However, the $l_1$-reconstruction is far from the information-theoretic limit \cite{Kabashima2009}.

If the vector $\textbf{x}$ is \emph{independent and identically distributed} (\emph{i.i.d.}) with known marginal distribution, and the noise $\textbf{n}$ is \emph{i.i.d.} Gaussian with known variance, the maximum a-posterior probability Bayesian estimation provides a minimum \emph{mean-square-error} (MSE) reconstruction, but the computational complexity will be extraordinarily unacceptable. Hence, from a belief-propagation perspective, a low-complexity iterative approximate Bayesian algorithm, named \emph{approximate message passing} (AMP) algorithm, is formulated \cite{Donoho2009, Rangan2010}. In \cite{Ma20151, CKWen20161, CKWen20162}, orthogonal measurement matrices (e.g. \emph{discrete Fourier transform} (DFT) matrices) are utilized to reduce the computational complexity and storage memory, and improve the convergence speed of the sparse vector recovery algorithms. Recently, a novel orthogonal AMP is proposed for a wide range of sensing matrices, including ill-conditioned matrices, partial orthogonal matrices, and general unitarily-invariant matrices \cite{Ma2017}. For the Gaussian-mixed vector with unknown sparsity, mean, and variance, and the noise as Gaussian with unknown variance, an \emph{expectation-maximization Gaussian-mixture AMP} (EM-GM-AMP) is designed \cite{Schniter2013}. However, all above works focus on the MSE of the value recovery while ignoring the sparsity requirement. The work in \cite{Tulino2013} focuses on the support (non-zero position) recovery rather than the MSE of the sparse vector recovery. Recently, a LSE-MP iterative algorithm is proposed for both support and value recovery \cite{Huang2016}. However, its computational complexity is high due to the need to perform matrix inversion in each iteration.

In this article, by using the knowledge of message passing \cite{Loeliger2006, Lei2015, Lei2016, Lei20162}, a low-complexity \emph{Bernoulli-Gaussian MP} (BGMP) algorithm considering both the value recovery and the support recovery is proposed, in which Bernoulli messages (for the value reconstruction) and Gaussian messages (for the support reconstruction) are jointly processed and assist each other iteratively. Our numerical results show that the proposed BGMP algorithm not only has a limit-approaching MSE in the value recovery, but also obtains an excellent SER performance in the support recovery.

\section{Problem Formulation}
%Consider the sparse vector recovery problem for the following noisy linear model
%\begin{equation}
%\mathbf{y}=\mathbf{H}\mathbf{x}+\mathbf{n},
%\end{equation}
%where $\mathbf{x}=[x_1,\cdots,x_{K}]^T$ is a sparse vector to be recovered, $\mathbf{H}\in {\mathbb{R}}^{M \times K}$ is a given measurement matrix, $\mathbf{n}\sim\mathcal{N}^{M} (0,\sigma_n^2)$ a vector of independent \emph{additive white Gaussian noise} (AWGN), and $\mathbf{y}\in {\mathbb{R}}^{M \times 1}$ a noisy measurement vector.
In this paper, we consider that the entries of  $\mathbf{x}$ are \emph{i.i.d.} and follows the Bernoulli-Gaussian distribution \cite{Ma20151}:
\begin{equation}\label{sp_mod}
x_k\sim\left\{ \begin{array}{l}
0, \qquad\qquad\quad\; \mathrm{probability} = 1 - \lambda,\\
\mathcal{N}(0,{\lambda ^{ - 1}}), \quad\;\: \mathrm{probability} = \lambda,\vspace{-0.1cm}
\end{array} \right.
\end{equation}
where $k\in \mathcal{K}$, $\mathcal{K}=\{1,\cdots,K\}$. In (\ref{sp_mod}), without losing any generality, the variance of ${x}_k$ is normalized to 1.

In this work, we try to recover the sparse vector, including positions of the zero components and values of the non-zero components, via \emph{message passing algorithm} (MPA). It is well known that there are a number of MPAs for the recovery of Gaussian or Bernoulli distributed $\mathbf{x}$, because the message update rules of Gaussian or Bernoulli random variables can be easily derived. However, to the best of the authors' knowledge, MPA for the recovery of Bernoulli-Gaussian distributed $\mathbf{x}$ is far from solved because of its complex message structure. To simplify the update of Bernoulli-Gaussian messages, we treat the Bernoulli-Gaussian random vector as a componentwise product of Bernoulli random vector $\mathbf{b}\sim \mathcal{B}^K(1,\lambda)$ and Gaussian random vector $\mathbf{g}\sim \mathcal{N}^K(0,\lambda^{-1})$, where $\mathbf{g}$ and $\mathbf{b}$ are independent with each other, i.e.\vspace{-0.1cm}
\begin{equation}
\mathbf{x}= \mathbf{g}\circ\mathbf{b},\vspace{-0cm}
\end{equation}
where $\mathbf{g}=[g_1,\cdots,g_{K}]^T$, $\mathbf{b}=[b_1,\cdots,b_{K}]^T$, and $\mathbf{x}=[g_1b_1,\cdots,g_{K}b_{K}]^T$. Therefore, the recovery of the sparse vector $\mathbf{x}$ is decomposed into recoveries of $\mathbf{g}$ and $\mathbf{b}$, which denote the value recovery and support recovery respectively.

\section{Bernoulli-Gaussian Message Passing}
\begin{figure}[t]
  \centering
  \includegraphics[width=8.5cm]{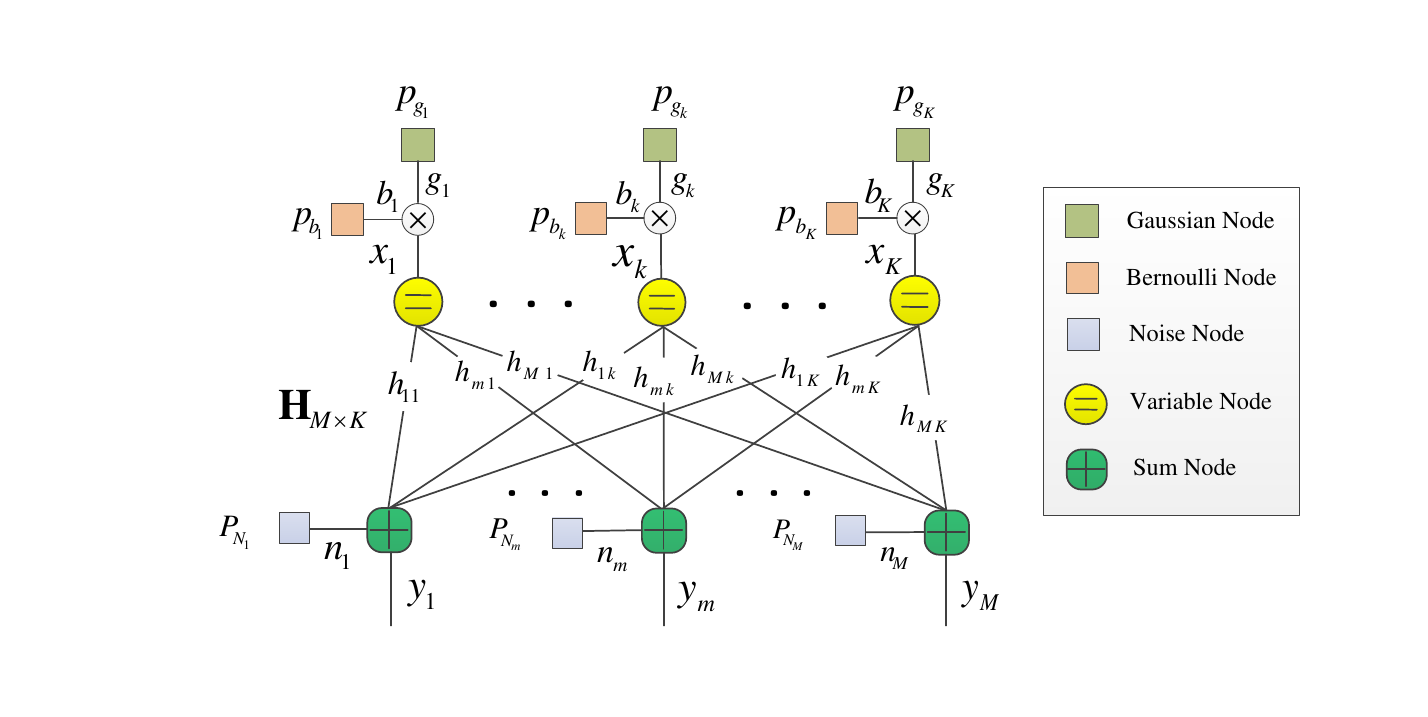}\\
  \caption{Factor graph of BGMP for sparse vector recovery.}\label{f1}
\end{figure}

In this section, we proposed a novel MPA, which jointly estimates $\mathbf{b}$ and $\mathbf{g}$, for the sparse vector recovery. Since the proposed algorithm updates both Bernuolli and Gaussian messages in the process, we call it BGMP algorithm. As shown in Fig. \ref{f1}, the BGMP is based on a pairwise factor graph, which consists of variable nodes, sum nodes, constraint nodes, and the corresponding edges. Message update in BGMP algorithm is similar to that of the \emph{belief propagation} (BP) decoding process of LDPC code, in which extrinsic messages are updated on the edges of the factor graph. Similar to distributed algorithms \cite{Duan2017}, the complexity of BGMP is very low since it decomposes the overall processing into many low-complexity calculations on the factor graph that can be executed in parallel. Apart from their similarity, there also exist differences between BGMP and BP or \emph{Gaussian message passing} (GMP) \cite{Lei2015, Lei2016, Lei20162}. One is that BGMP updates both Gaussian and Bernoulli messages on the factor graph, while the BP deals with only Bernoulli messages and GMP with Gaussian messages. The other is the different message update functions on the factor graph. The detailed message updating rules are derived as follows.
\begin{figure*}[t]\vspace{-0.3cm}
  \centering
  \includegraphics[width=13cm]{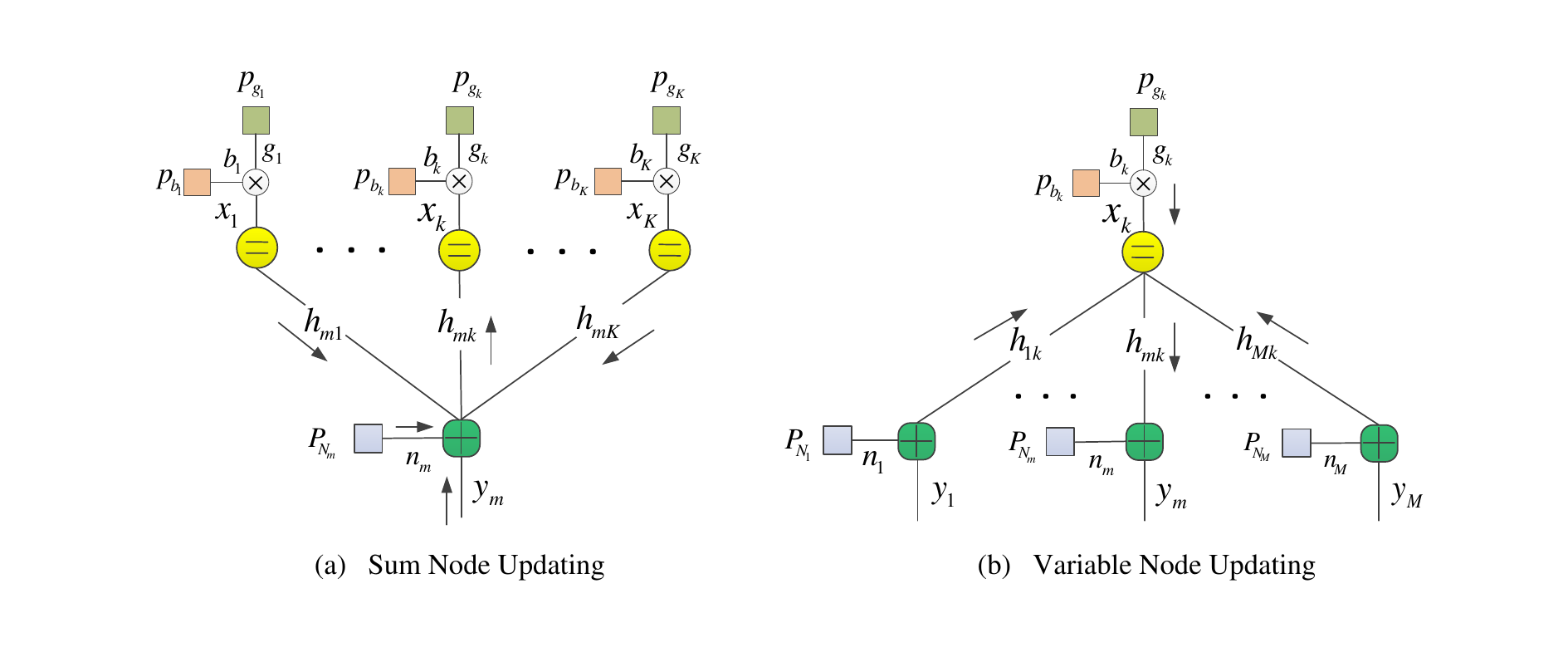}\\
  \caption{Message update at sum nodes and variable nodes. For the Bernoulli-Gaussian signal, the mean and variance of a Gaussian distribution, and the non-zero probability of a Bernoulli distribution are passing on each edge. An extrinsic message is updated on each edge via the messages on the other edges that are connected with the same node. }\label{f2}
\end{figure*}
\subsection{Bernoulli-Gaussian Message Update at Sum Node}
In the left subfigure of Fig. \ref{f2}, each sum node is treated as a multiple-access process, and we derive the Bernoulli-Gaussian message update at the \emph{variable node} (VN). Firstly, the received $y_m$ at the $m$-th SN can be rewritten to
\begin{eqnarray}
y_m \!\!\!&=&\!\!\! h_{mk}x_k + \sum\limits_{i\in \mathcal{K}/k} {h_{mi}x_{i}} + n_m\nonumber\\
    &=&\!\!\! h_{mk}g_kb_k + \underbrace{\sum\limits_{i\in \mathcal{K}/k} {h_{mi}g_{i}b_i} + n_m}_{n_{mk}^*},
\end{eqnarray}
where $m\in\mathcal{M}$, $\mathcal{M}=\{1,\cdots, M\}$, $h_{mk}$ is an element of $\mathbf{H}$, and $i\in \mathcal{K}/k$ denotes $i\in \mathcal{K} \;\mathrm{and}\; i\neq k$. As the $\{g_i,\;b_i\}, i\in \mathcal{K}$ are independent with each other, we can approximate $\sum\limits_{i\in \mathcal{K}/k} {h_{mi}g_{i}b_i} + n_m$ as an equivalent Gaussian noise $n_{mk}^*\sim\mathcal{N}\big(u_{mk}^*,v_{mk}^*\big)$ based on \emph{central limit theorem}:\vspace{-0.1cm}
\begin{equation}
y_m = h_{mk}g_kb_k + n_{mk}^*.\vspace{-0.1cm}
\end{equation}
Let $u_{k\to m}^{v}(\tau)$ and $v_{k\to m}^{v}(\tau)$ denote the mean and variance of the Gaussian variable $g_k$ passing from $k$-th VN to $m$-th SN in $\tau$-th iteration. Similarly, $p_{k\to m}^v(\tau)$ denotes the non-zero possibility of the Bernoulli variable $b_k$ passing from $k$th VN to $m$th SN. In $\tau$-th iteration, the mean and variance of the noise $n_{mk}^*$ can be derived directly:
\begin{equation}\label{equ_noi}
\left\{ \!\!\!\!\begin{array}{l}
u_{mk}^*(\!\tau\!) \!\!= \!\!\!\!\sum\limits_{i \in {\cal K}/k} \!\!\!\!{{h_{mi}}p_{i \to m}^v(\!\tau\!)u_{i \to m}^v}(\!\tau\!),\\
v_{mk}^*\!(\!\tau\!)\!\! =\!\! \!\!\!\sum\limits_{i \in\! {\cal K}\!/\!k} \!\!\!{h_{mi}^2p_{i \to m}^v\!(\!\tau\!)[ {v_{i \to m}^v(\!\tau\!) \!+\! (1\!\! -\! p_{i \to m}^v(\!\tau\!)\!)u_{i \to m}^{{v^2}}(\!\tau\!)} ] \!\!+\! \sigma _n^2},
\end{array} \right.\vspace{-0.1cm}
\end{equation}
where $m\in\mathcal{M}$ and $k\in\mathcal{K}$. Let ${\bf{V}}_v(\tau)$, $\mathbf{U}_v(\tau)$ and $\mathbf{P}_v(\tau)$ be the matrixes containing the elements $v_{k\to m}^v(\tau)$, $u_{k\to m}^v(\tau)$ and $p_{k\to m}^v(\tau)$, $\forall m\in\mathcal{M}$, $\forall k\in\mathcal{K}$, respectively. Then, ${\bf{V}}_v(0)$ is initialized to $+\boldsymbol{\infty}$, ${\bf{U}}_v(0)$ to $\mathbf{0}$, and ${\bf{P}}_v(0)$ to $0.5*\mathbf{1}$.
\subsubsection{Gaussian message update for $\mathbf{g}$}
Let $u_{m\to k}^{s}(\tau)$ and $v_{m\to k}^{s}(\tau)$ denote the mean and variance of $g_k$, and $p_{m\to k}^s(\tau)$ the non-zero possibility of $b_k$, passing from $m$-th SN to $k$-th VN. Then, the message update of $g_k$ at $m$-th SN for $k$-th VN is derived by $[u_{mk}^{*}(\tau), v_{mk}^{*}(\tau)]$:
\begin{equation}\label{sum_gau}
\left\{ \!\!\!\!\!\!\!{\begin{array}{*{20}{l}}
\begin{array}{l}
u_{m \to k}^s(\tau) \mathop  = \limits^{(a)}  \mathrm{E}\left[ {{g_k}|y_m ,u_{mk}^*(\tau),v_{mk}^*(\tau),{b_k} = 1} \right]\vspace{0.1cm}\\
\quad\quad \quad \;\;\;  = h_{mk}^{ - 1}\big({y_m} - u_{mk}^*(\tau)\big),
\end{array}\\\vspace{-0.2cm}
\\
\begin{array}{l}
v_{m \to k}^s(\tau) \mathop  = \limits^{(b)} {\rm{Var}}\left[ {{g_k}|y_mu_{mk}^*(\tau),v_{mk}^*(\tau),{b_k} = 1} \right]\vspace{0.1cm}\\
\quad\quad \quad \;\;\; = h_{mk}^{ - 2}v_{mk}^*(\tau),
\end{array}
\end{array}} \right.
\end{equation}
where $\mathrm{E}(a|d)$ and $\mathrm{Var}(a|d)$ denote the conditional expectation and variance of variable $a$ given $d$, respectively. The equations (a) and (b) in (\ref{sum_gau}) are obtained by the fact that there is no information for $g_k$ given $b_k=0$. %where ${\bf{p}}_m^v(\tau)\!\!=\![p_{1\to m}^v(\tau),\!\cdots\!,p_{K\!\to m}^v(\tau)]^T$, ${\bf{u}}_m^v(\tau)\!\!=\![u_{1\to m}^v(\tau),\!\cdots\!,u_{K\!\to m}^v(\tau)]^T$, and ${\bf{v}}_m^v(\tau)=[v_{1\to m}^v(\tau),\cdots,v_{K\to m}^v(\tau)]^T$. In addition, ${\bf{u}}_{m,\sim k}^v(\tau)$, ${\bf{v}}_{m,\sim k}^v(\tau)$ and ${\bf{p}}_{m,\sim k}^v(\tau)$ are obtained from ${\bf{u}}_{m}^v(\tau)$, ${\bf{v}}_{m}^v(\tau)$ and ${\bf{p}}_{m}^v(\tau)$ by excluding their $k$th entries ${{u}}_{mk}^v(\tau)$, ${{v}}_{mk}^v(\tau)$ and ${{p}}_{mk}^v(\tau)$ respectively.

\subsubsection{Bernoulli message update for $\mathbf{b}$} Similarly, the message update of $b_k$ at $m$-th SN for $k$-th VN is also derived by $[u_{mk}^{*}(\tau), v_{mk}^{*}(\tau)]$:\vspace{-0.1cm}
\begin{eqnarray}\label{sum_Ber}
p_{\!m \to k}^s \!(\!\tau\!)
\!\!\!\!\!&=&\!\!\!\!\!\! \left[1+\frac{{P(y_m |{b_k} = 0,u_{mk}^*(\tau),v_{mk}^*(\tau))}}{{P(y_m|{b_k} \!=\! 1 ,u_{mk}^*(\tau),v_{mk}^*(\tau)) }}\right]^{-1}\mathop {}\limits_{\mathop {\mathop {}\limits_{} } } \nonumber\\
&= & \!\!\!\!\!\!\left[\!1\!+\!\frac{{P(y_m = n_{mk}^*|u_{mk}^*(\tau),v_{mk}^*(\tau))}}{{P\big(y_m \!=\! h_{mk}g_k + n_{mk}^*|u_{mk}^*(\tau),v_{mk}^*(\tau)\big)}} \right]^{-1}\mathop {}\limits_{\mathop {\mathop {}\limits_{} } }\nonumber\\
\!& =& \!\!\!\!\!\!\!\left[\!1\!\!+\!\!\tfrac{ f\big({y_m}|u_{mk}^*(\!\tau\!),v_{mk}^*(\!\tau\!)\big)}{{f\big({y_m}|u_{mk}^*\!(\!\tau\!) + {h_{mk}}\!u_{k \to m}^v(\!\tau\!),v_{mk}^*(\!\tau\!) + {h_{m\! k}^2}\!v_{k \to m}^v(\!\tau\!)\big) }}\!\right]^{\!-\!1}
%&=& \!\!\!\!\!\!\left[\!1\!\! +\! \!\sqrt{\!1\!\!+\!\tfrac{v_{k \to m}^v(\!\tau\!)}{v_{m \to k}^s(\tau)}} \mathrm{Exp}\!\Big({\!-\tfrac{v_{k \to m}^v\!(\!\tau\!)u_{m \to k}^{s^2}(\!\tau\!)\! +\! u_{k\to m}^v(\!\tau\!)v_{m \to k}^s(\!\tau\!) \big(\!2u_{m \to k}^s(\!\tau\!)\!-\!u_{k\to m}^v(\!\tau\!)\!\big)}{2v_{m \to k}^s(\!\tau\!)\big(v_{m \to k}^s(\!\tau\!)+v_{k \to m}^v(\!\tau\!)\big)}}\!\Big) \!\!\right]^{\!\!-1}\!\!\!\!\!\!\!,\quad\,
\end{eqnarray}
where $f(x|u,v)$ is a \emph{probability density function} (PDF) of a Gaussian distribution $\mathcal{N}(u,v)$, i.e.,\vspace{-0.1cm}
\begin{equation}
f(x|u,v)=\frac{1}{\sqrt{2\pi v}} e^{-\frac{(x-u)^2}{2v}}.
\end{equation}

\subsection{Bernoulli-Gaussian Message Update at Variable Node}
In the right subfigure of Fig. \ref{f2}, each variable node is treated as a broadcast process, and we derive the Bernoulli-Gaussian message update at the VN. According to the message combination rule \cite{ Loeliger2006, Lei2016}, the messages of the same variable are combined by a normalized product of the input PDFs. As the $\mathbf{g}$ and $\mathbf{b}$ are \emph{i.i.d.}, and independent each other, we update the messages for $\{g_i , i\in \mathcal{K}\}$ and $\{b_i , i\in \mathcal{K}\}$ independently.
\subsubsection{Gaussian message update for $\mathbf{g}$}
Let $\bar{\mathbf{u}}=[\bar{u}_1,\cdots,\bar{u}_K]^T$ and $\bar{\mathbf{v}}=[\bar{v}_1,\cdots,\bar{v}_K]^T$ be the prior mean and variance of the Gaussian vector $\mathbf{g}$, $\bar{\mathbf{p}}=[\bar{p}_1,\cdots,\bar{p}_K]^T$ be the prior non-zero probability of the Bernoulli vector $\mathbf{b}$. Set $\mathcal{M}/m$ is obtained from set $\mathcal{M}$ by excluding the element $m$. Without loss of generality, we assume that $\bar{u}_i = 0$, $\bar{v}_i= \lambda^{-1}$ and $\bar{p}_i=\lambda$ for any $i\in \mathcal{K}$. The Gaussian message of $g_k$ at $k$-th VN for $m$-th SN is updated by the Gaussian messages from the SN set $\mathcal{M}/m$.
\begin{equation}\label{var_gau}
\left\{ \!\!\!\!\!{\begin{array}{*{20}{l}}
\begin{array}{l}
v_{k \to m}^v (\!\tau\!+\!1\!)= {\rm{Var}}\left[ {{g_k}|{\mathbf{v}}_{k,\sim m}^s}(\!\tau\!),\bar{{v}}_k \right]\\
\qquad\qquad\quad\mathop  = \limits^{(a)} [\lambda +\! \sum \limits_{j\in \mathcal{M}/m}{\!\!\!v_{j\to k}^{s^{-1}}(\!\tau\!)}]^{-1},
\end{array}\\\vspace{-0.2cm}
\\
\begin{array}{l}
u_{k \to m}^v (\!\tau\!+\!1\!) = \mathrm{E}\left[ {g_k}|{\mathbf{v}}_{k,\sim m}^s(\!\tau\!),{\mathbf{u}}_{k,\sim m}^s(\!\tau\!),\bar{{v}}_k,\bar{{u}}_k \right]  \\
\qquad\qquad\quad\;\mathop  = \limits^{(b)} v_{k\to m}^v(\!\tau\!) \!\!\sum \limits_{j\in \mathcal{M}/m}{\!\!\!v_{j\to k}^{s^{-1}}(\!\tau\!)}u_{j\to k}^{s}(\!\tau\!),
\end{array}
\end{array}} \right.
\end{equation}
where $m\in\mathcal{M}$, $k\in\mathcal{K}$, ${\bf{u}}_k^s(\!\tau\!)=[u_{1\to k}^s(\!\tau\!),\cdots,u_{M\to k}^s(\!\tau\!)]^T$, ${\bf{v}}_k^s(\!\tau\!)=[v_{1\to k}^s(\!\tau\!),\cdots,v_{M\to k}^s(\!\tau\!)]^T$, and ${\bf{u}}_{k,\sim m}^s(\!\tau\!)$ and ${\bf{v}}_{k,\sim m}^s(\!\tau\!)$ are obtained from ${\bf{u}}_{k}^s(\!\tau\!)$ and ${\bf{v}}_{k}^s(\!\tau\!)$ by excluding their $k$-th entries ${{u}}_{mk}^s(\!\tau\!)$ and ${{v}}_{mk}^s(\!\tau\!)$ respectively. Equations \emph{(a)} and \emph{(b)} are obtained by the combination of Gaussian PDFs \cite{ Loeliger2006, Lei2016}.
\subsubsection{Bernoulli message update for $\mathbf{b}$}
The Bernoulli message of $b_k$ at the $k$-th VN for the $m$th SN is derived by the Bernoulli messages from SN set $\mathcal{M}/m$.\vspace{-0.1cm}
\begin{eqnarray}\label{var_ber}
\!\!\!\!\!\!\!\!\!\!\!\! p_{k \to m}^v\!(\!\tau\!\!+\!\!1\!)\!\!\!\!\!&=&\!\!\!\!\! P(b_k|\mathbf{p}_{k,\sim m}^{s}(\!\tau\!), \bar{p}_k)\nonumber\\
&\mathop  = \limits^{(c)}& \!\!\!\!  \frac{{{\lambda}\!\!\!\mathop \Pi \limits_{j \in \mathcal{M}\!/\!m} \!\!\!p_{j \to k}^s(\!\tau\!)}}{{{\lambda}\!\!\!\mathop \Pi \limits_{j \!\in\! \mathcal{M}\!/\!m} \!\!\!p_{\!j \to k}^s(\!\tau\!) \!+\! (1 \!\!-\!\! \lambda)\!\!\!\mathop \Pi \limits_{j \!\in\! \mathcal{M}\!/\!m} \!\!\!(1 \!-\! p_{j \to k}^s(\!\tau\!))}},\vspace{-0.1cm}
\end{eqnarray}
where $m\in\mathcal{M}$, $k\in\mathcal{K}$, $\mathbf{p}_{k}^{s}(\!\tau\!)=[p_{1\to k}^s(\!\tau\!),\cdots,p_{M\to k}^s(\!\tau\!)]^T$, and ${\bf{p}}_{k,\sim m}^s(\!\tau\!)$ is obtained from ${\bf{p}}_{k}^s(\!\tau\!)$ by excluding the $k$-th entry ${{p}}_{mk}^s(\!\tau\!)$. Equation \emph{(c)} is derived by combination of Bernoulli PDFs \cite{ Loeliger2006}.

\subsection{Decision and Output of BGMP}
The BGMP algorithm iteratively performs the message update at the SNs and the VNs. When the MSE meets the requirement or the number of iterations reaches the limit, we output the $\hat{u}_{k}$ and $\hat{v}_{k}$ as the final estimate and deviation of $g_k$, and the non-zero probability $\hat{p}_k$ of $b_k$.
\begin{equation}\label{var_dec}
\!\!\left\{ \!\!\!\begin{array}{l}
\hat{v}_{k} =  \big(\lambda +\! \sum \limits_{m\in \mathcal{M}}{\!\!\!v_{m\to k}^{s^{-1}}}(\!\tau\!)\big)^{-1}\!\!\!\!,\mathop {}\limits_{\mathop { }\limits_{\mathop { }\limits_{}}} \\
\hat{u}_{k } =  \hat{v}_{k}  \!\!\sum \limits_{m\in \mathcal{M}}{\!v_{m\to k}^{s^{-1}}}(\!\tau\!)u_{m\to k}^{s}(\!\tau\!),\mathop {}\limits_{\mathop { }\limits_{\mathop { }\limits_{\mathop { }}}} \\
\hat{p}_{k} =   \dfrac{{{\lambda}\!\!\!\mathop \Pi \limits_{m \in \mathcal{M}} \!\!\!p_{m \to k}^s}(\!\tau\!)}{{{\lambda}\!\!\!\mathop \Pi \limits_{m \in \mathcal{M}} \!\!\!p_{m \to k}^s(\!\tau\!) +(1 \!-\! \lambda)\!\!\!\mathop \Pi \limits_{m \in \mathcal{M}} \!\!\!(1 \!-\! p_{m \to k}^s(\!\tau\!))}},
\end{array} \right.
\end{equation}
where $k\in\mathcal{K}$. Then, final estimate of $\mathbf{b}$ is given by
\begin{equation}
\hat{b}_k=\left\{ \begin{array}{l}
1, \;\;\mathrm{if}\;\; \hat{p}_k\geq0.5, \\
0, \;\;\mathrm{if}\;\; \hat{p}_k<0.5,
\end{array} \right.
\end{equation}
for $k\in\mathcal{K}$. Let $\hat{\mathbf{u}}=[\hat{{u}}_1,\cdots,\hat{{u}}_K]^T$, $\hat{\mathbf{p}}=[\hat{p}_1,\cdots,\hat{p}_K]$, and $\hat{\mathbf{b}}=[\hat{{b}}_1,\cdots,\hat{{b}}_K]^T$. The final a-posterior estimate of the sparse vector $\mathbf{x}$ is
\begin{equation}
\hat{\mathbf{x}}=\hat{\mathbf{p}}\circ\hat{\mathbf{u}}\circ\hat{\mathbf{b}},
\end{equation}
and its mean square error (MSE) is
\begin{equation}
\mathbf{mse}_{\hat{\mathbf{x}}}=\hat{\mathbf{p}}\circ\big(\hat{\mathbf{v}} +(1-\hat{\mathbf{p}})\circ\hat{\mathbf{u}}^{(2)}\big).
\end{equation}
\subsection{LLR-based BGMP}
The message updates for the Bernoulli vector always overflow due to the probability multiplications. To avoid the overflow, the following \emph{log-likelihood ratios} (LLRs) are utilized to replace the non-zero probabilities in BGMP.
\begin{equation*}
l_{m\to k}^s(\!\tau\!)\!=\!\mathcal{L}\big(p_{m\to k}^s(\!\tau\!)\big),\;l_{k\to m}^v(\!\tau\!)\!=\! \mathcal{L}\big(p_{k\to m}^v(\!\tau\!)\big) ,\; \bar{l}_k\!=\!\mathcal{L}\big(\bar{p}_k\big),
\end{equation*}
for any $k\in {\mathcal{K}}$ and $m\in {\mathcal{M}}$, where $\mathcal{L}(p)=-\log(1-p^{-1})$. Then, the LLR-based BGMP algorithm is rewritten as follows.
%\newpage
\subsubsection{Message Update at SN} The Bernoulli-Gaussian message update at SN is the same as that in (\ref{sum_gau}) and (\ref{sum_Ber}), with $p_{k \to m}^v(\!\tau\!)=1/(1+e^{-l_{k \to m}^v(\!\tau\!)})$, $m\in\mathcal{M}$ and $k\in\mathcal{K}$. Let ${\bf{L}}_v(\tau)$ be a matrix containing $l_{k\to m}^{v}(\tau)$, $m\in\mathcal{M}$ and $k\in\mathcal{K}$. Then,  ${\bf{L}}_v(0)$ is initialized as $\mathbf{0}$.

\subsubsection{LLR Update at VN} The Bernoulli message update at $k$-th VN for $m$-th SN is rewritten to
\begin{equation}
l_{k \to m}^v(\tau\!+\!1) = \bar{l}_k + \!\!\sum \limits_{j\in \mathcal{M}/m}{\!\!{l}_{j\to k}^s}(\!\tau\!),\;\; m\in\mathcal{M}, k\in\mathcal{K}.
\end{equation}

\subsubsection{LLR Output}When the MSE of the BGMP meets the requirement or the number of iterations reaches the limit, we output the final LLR $\hat{l}_k$ of the Bernoulli variable $b_k$.
\begin{equation}
\hat{l}_{k} =   \bar{l}_k + \!\sum \limits_{j\in \mathcal{M}}{\!{l}_{j\to k}^s}(\!\tau\!),\;\;k\in\mathcal{K}.
\end{equation}
Then, final estimate of $\mathbf{b}$ is given by
\begin{equation}\label{indi_fun}
\hat{b}_k=\left\{ \begin{array}{l}
1, \;\;\mathrm{if}\;\; \hat{l}_k\geq0 \\
0, \;\;\mathrm{if}\;\; \hat{l}_k<0
\end{array} \right.
\end{equation}
for $k\in\mathcal{K}$. Let $\hat{\mathbf{l}}=[\hat{l}_1,\cdots,\hat{l}_K]^T$, and $\hat{p}_k=(1+e^{-\hat{l}_k})^{-1}$. Then, $\mathbf{b}$ is recovered by an indicate function, i.e., $\hat{\mathbf{b}}=\mathcal{I}_{\;\hat{\mathbf{l}}}$. The final a-posterior estimate of the sparse vector $\mathbf{x}$ is
$\hat{\mathbf{x}}=\hat{\mathbf{p}}\circ\hat{\mathbf{u}}\circ\hat{\mathbf{b}}$,
and its mean square error $\mathbf{mse}_{\hat{\mathbf{x}}}= \hat{\mathbf{p}}\circ\big(\hat{\mathbf{v}} +(1-\hat{\mathbf{p}})\circ\hat{\mathbf{u}}^{(2)}\big)$.

\subsection{BGMP in Matrix Form}
\emph{Note:} We let $\mathbf{A}_{M\times N}\circ\mathbf{B}_{M\!\times\! N}=\left[ a_{ij}b_{ij}\right]_{M\times N}$, $\frac{\mathbf{A}_{M\!\times\! N}}{\mathbf{B}_{M\!\times\! N}} = \left[ a_{ij}/b_{ij}\right]_{M\!\times\! N}$, $\mathrm{Exp}(\mathbf{A}_{M\!\times\! N})=[e^{a_{ij}}]_{M\times N}$, $\mathcal{D}\{\mathbf{A}_{N\!\times\! N}\}=\left[a_{ii}\right]_{N\!\times\! 1}$, $\mathbf{1}_{M\!\times\! N} = \left[1\right]_{M\!\times\! N}$, and $\mathbf{A}^{(k)}_{M\!\times\! N}=\left[a^k_{ij}\right]_{M\!\times\! N}$. Assume $\bar{\mathbf{l}}=[\bar{l}_1,\cdots,\bar{l}_K]^T$, $\mathbf{U}_{s}(\tau)\!=\!\left[u_{m\to k}^s(\tau)\right]_{M\!\times\! K}$, $\mathbf{V}_{s}(\tau)\!=\!\left[v_{m\to k}^s(\tau)\right]_{M\!\times\! K}$, $\mathbf{L}_s(\tau)\!=\![l_{m\to k}^s(\tau)]_{M\!\times\! K}$, $\mathbf{U}_{v}(t)\!=\!\left[u_{k\to m}^v(\tau)\right]_{K\!\times\! M}$, $\mathbf{V}_{v}(\tau)=\left[v_{k\to m}^v(\tau)\right]_{K\!\times\! M}$, $\mathbf{L}_v(\tau)=[l_{k\to m}^v(\tau)]_{K\!\times\! M}$, and $\mathbf{P}_v(\tau)=\left[p_{k\to m}^v(\tau)\right]_{K\!\times\! M}$. Algorithm 1 shows the detailed process of matrix-form BGMP.

\begin{algorithm}
\caption{BGMP Algorithm}
\begin{algorithmic}[1]
\State {\small{\textbf{Input:} {{$\mathbf{H}$, $\mathbf{V}_{\!\textbf{\emph{x}}}$, $\sigma^2_n$, $\lambda\!\!\in\!\! (0,1)$, $\epsilon\!\!>\!\!0$, $N_{ite}$}} and {{$\mathbf{H}^{(2)}$}}, $\mathbf{H}^{(\!-1)}$, $\mathbf{H}^{(\!-2)}$.
\State \textbf{Initialization:} $\tau=0$, $\mathbf{U}_v(\!0\!)=\mathbf{0}$, $\mathbf{V}_v(\!0\!)\!=\!+\boldsymbol{\infty}$, and $\mathbf{L}_v(\!0\!)\!=\!\mathbf{\!0}$.
\State \textbf{Do}
\State { $\mathbf{P}_v(\!\tau\!)\!=\!\big(\mathbf{1}_{K\!\times\! M}\!+e^{\!-\mathbf{L}_v(\!\tau\!)}\big)^{(\!-\!1\!)}$, $\widetilde{\mathbf{U}}^*(\tau) \!=\! \mathbf{H}^T\circ\mathbf{P}_v(\tau)\circ\mathbf{U}_v(\tau)$, \\ $\;\,\widetilde{\mathbf{V}}^*(\tau) \!\!=\!\! \mathbf{H}^{(2)^T}\!\!\circ\mathbf{P}_v(\!\tau\!)\!\circ\!\big( \mathbf{V}_v(\!\tau\!)+ (\mathbf{1}_{K\times M}\!-\!\mathbf{P}_v(\!\tau\!))\circ \mathbf{U}_v^{(2)}(\!\tau\!) \big)$.
\vspace{0.1cm}
\State  \vspace{-0.25cm}\[
\!\!\!\!\!\!\!\!\!\begin{array}{l}
\left[ \!\!\!\!\begin{array}{l}
\mathbf{U}^*\!(\!\tau\!)\\
\mathbf{V}^*\!(\!\tau\!)
\end{array} \!\!\!\!\right] \!\!\!=\!\!\! \left[ \!\!\!\begin{array}{c}
\mathcal{D}\{ \mathbf{1}_{M\times K}\cdot\widetilde{\mathbf{U}}^*(\tau)\} \mathop {}\limits_{\mathop { }}\\
\sigma_n^2\!\!\cdot\!\!\mathbf{1}_{\!M\!\times \!1}\!\!+\!\mathcal{D}\{ \mathbf{1}_{\!M\!\times \!K}\!\cdot\!\! \widetilde{\mathbf{V}}^*(\!\tau\!) \}
\end{array}\!\!\! \right]\! \!\cdot\!\! {\mathbf{1}_{1 \times K}}
\!-\!\left[ \!\!\!\begin{array}{c}
\widetilde{\mathbf{U}}^{*^T}\!(\!\tau\!) \mathop {}\limits_{\mathop { }}\\
 \widetilde{\mathbf{V}}^{*^T}\!\!(\!\tau\!)
\end{array}\!\!\!\! \right],
\end{array}\]}}
\State\vspace{-0.25cm}\[ \!\!\!\!\!\!\!\!\begin{array}{c}
\left[\!\!\!\! \begin{array}{c}
{{\rm{\mathbf{U}}}_{\!s}}(\!\tau\!)\\
{\mathbf{V}_{\!s}}(\!\tau\!)\\
\mathbf{L}_{\!s}(\!\tau\!)
\end{array} \!\!\!\!\right]\!\! \!=\!\!\! \left[\! \!\!\!\begin{array}{c}
\mathbf{H}^{(-1)}\circ\left({{\mathbf{y}}}\cdot{\mathbf{1}_{1 \times K}} - \mathbf{U}^*(\tau)\right) \mathop {}\limits_{\mathop { }}\\
\mathbf{H}^{(-2)}\circ\mathbf{V}^*(\tau)\mathop {}\limits_{\mathop { }}\\
{{\!-\!\frac{1}{2}}}\!\log{\!\big[\!{1\!\!+\!\! \tfrac{\mathbf{V}_v^T\!\!(\!\tau\!)} {{\mathbf{V}_{\!s}}\!(\!\tau\!)}\!\big]} \!\!+\!\! \tfrac{\mathbf{V}^T_v\!\!(\!\tau\!) \circ\mathbf{V}^{(\!-\!1\!)}_s\!(\!\tau\!)\!\circ\!{\mathbf{U}_{\!s}}\!(\!\tau\!)^{\!(\!2\!)} \!- \mathbf{U}^T_v\!(\!\tau\!)\!\circ\! \left(\!2{\mathbf{U}_{\!s}}\!(\!\tau\!)-\mathbf{U}^T_v\!(\!\tau\!)\right)} {2\big(\mathbf{V}_s(\!\tau\!)+ \mathbf{V}^T_v(\!\tau\!)\big)}}
\end{array} \!\!\!\!\!\right]\!\!\!,
\end{array}\]
\State \vspace{-0.1cm}\[\!\!\!\!\!\!\!\!\! \begin{array}{c}
\left[\!\!\!\!\! \begin{array}{c}
{\mathbf{V}_{\!v}}\!(\!\tau\!\!+\!\!1\!)\\
{{\rm{\mathbf{U}}}_{\!v}}\!(\!\tau\!\!+\!\!1\!)\\
\mathbf{L}_{\!v}\!(\!\tau\!\!+\!\!1\!)
\end{array} \!\!\!\!\!\right]\!\! \!=\!\!\! \left[\!\! \!\!\!\begin{array}{c}
\big[\lambda\!\!\cdot\!\!\mathbf{1}_{K\!\times\! M} \!+\! \mathcal{D}\{\mathbf{1}_{\!K\!\times\! M}\!\!\cdot\!\!\mathbf{V}_{\!s}^{(\!-\!1\!)}\!(\!\tau\!)\}\!\!\cdot\!\! \mathbf{1}_{\!1\times\! M}\!-\!\!\mathbf{V}_{\!s}^{(\!-\!1\!)^T}\!\!\!(\!\tau\!)\!\big]^{\!(\!-\!1\!)} \mathop {}\limits_{\mathop { }}\\
{{\rm{\mathbf{V}}}_{\!\!v}}\!(\!\tau\!\!+\!\!1\!) \!\!\circ\!\!\! \Big[ \!\mathcal{D}\!\{\!\mathbf{1}_{\!K\!\times\! M}\!\!\cdot\!\!\big(\mathbf{V}_{s}^{\!(\!-\!1\!)}\!(\!\tau\!)\!\!\circ\!\! \mathbf{U}_{\!s}\!(\!\tau\!) \!\big)\!\}\!\!\cdot\!\!\mathbf{1}_{\!1\!\times\! M}\!\!-\!\! \mathbf{V}_{s}^{\!(\!-\!1\!)^{\!T}}\!\!\!(\!\tau\!)\!\!\circ\!\! \mathbf{U}_{s}^{\!T}\!(\!\tau\!)  \!\Big]\mathop {}\limits_{\mathop { }} \\
 \big[\mathcal{D}\{\mathbf{1}_{\!K\!\times\! M}\cdot\mathbf{L}_{s}(\tau)\}+\bar{\mathbf{l}}\,\big]\cdot\mathbf{1}_{1\times M}-\mathbf{L}_{s}^{^T}(\tau) \mathop {}\limits_{\mathop { }}
\end{array} \!\!\!\!\!\!\right]\!\!\!,
\end{array}\]
\vspace{0cm}
\State \; $\tau=\tau+1$.
\vspace{0.15cm}
\State \textbf{While} {\small{$\big(\! (|\mathbf{U}_{\!v}\!(\!\tau\!+\!1\!)\!-\!\!\mathbf{U}_{\!v}{\!(\!\tau\!)}|\!<\!\epsilon \& |\mathbf{L}_{v}\!(\!\tau\!+\!1)\!-\!\mathbf{L}_{v}\!{(\!\tau\!)}|\!<\!\epsilon) \;{\textbf{or}}\; \tau\!\leq\! N_{ite}\big)$}}
\State  \vspace{-0.25cm}{{ \[\!\!\!\!\begin{array}{c}
\left[\!\!\! \begin{array}{c}
\hat{\mathbf{v}}\\
\hat{\mathbf{u}}\\
\hat{\mathbf{l}}
\end{array} \!\!\!\right]\!\! \!=\!\!\! \left[ \!\!\!\begin{array}{c}
\big[\lambda\cdot\mathbf{1}_{K\times 1} + \mathcal{D}\{\mathbf{1}_{K\times M}\cdot\mathbf{V}_{s}^{(-1)}(\tau)\}\big]^{(-1)} \mathop {}\limits_{\mathop { }}\\
\hat{\mathbf{v}} \circ \mathcal{D}\{\mathbf{1}_{K\times M}\!\cdot\!\!\big(\mathbf{V}_{s}^{(-1)}(\!\tau\!)\!\circ\! \mathbf{U}_{s}(\!\tau\!) \big)\} \mathop {}\limits_{\mathop { }} \\
 \bar{\mathbf{l}} + \mathcal{D}\{\mathbf{1}_{K\times M}\cdot\mathbf{L}_{s}(\tau)\} \mathop {}\limits_{\mathop { }}
\end{array} \!\!\!\right]\!\!.
\end{array}\qquad\qquad\qquad\qquad\qquad\qquad\qquad\qquad\qquad\qquad\;\]}}
\State  $\hat{\mathbf{p}}=(\mathbf{1}_{K\times 1}+e^{-\hat{\mathbf{l}}})^{(-1)}$, $\hat{\mathbf{b}}=\mathcal{I}_{\;\hat{\mathbf{l}}}$, $\hat{\mathbf{x}}=\hat{\mathbf{p}}\circ\hat{\mathbf{u}}\circ\hat{\mathbf{b}}$, and \qquad$\mathbf{mse}_{\hat{\mathbf{x}}}=\hat{\mathbf{p}}\circ\big(\hat{\mathbf{v}} +(1-\hat{\mathbf{p}})\circ\hat{\mathbf{u}}^{(2)}\big)$.
\State \textbf{Output:} $\hat{\mathbf{b}}$, ${\hat{\textbf{\emph{x}}}}$,
and $\mathbf{mse}_{\hat{\mathbf{x}}}$ }.
\end{algorithmic}
\end{algorithm}\vspace{-0.2cm}

\subsection{Approximated Bernoulli message update at SN}
Due to the fact that $v_{k\to m}^v(\tau)<<v_{m\to k}^s(\tau)$, Bernoulli message update at SN (\ref{sum_Ber}) can be approximated to
\begin{equation*}
\!\!p_{\!m \to k}^s\! (\!\tau\!)\!\!= \!\!\left[\!1 \!\!+\!\! \mathrm{Exp}\big(\!\!-\!\!z_{mk}^s(\tau)\big( \!u_{\!k\to m}^v\!(\!\tau\!) \!+\! 0.5z_{mk}^s(\tau)v_{\!k \to m}^v\!(\!\tau\!) \!\big)\big)\right]^{\!-\!1}\!\!\!\!,
\end{equation*}
where $z_{mk}^s(\tau)=v_{\!m \to k}^{s^{-1}}\!(\!\tau\!)u_{\!m \to k}^{s}\!(\!\tau\!)$, and its LLR to
\begin{equation*}\label{ap_sum_ber}
\!\!\!\mathbf{L}_s(\!\tau\!)\!=\!\!
\mathbf{Z}_s\!(\!\tau\!)\!\circ\!\!\left[  \!\mathbf{U}^T_v\!(\!\tau\!)\!+\! 0.5\mathbf{Z}_s\!(\!\tau\!)\!\circ\!\!\mathbf{V}_s^{(\!-\!1)}\!(\!\tau\!)\!\right]\!\!, \mathbf{Z}_s\!(\!\tau\!)\!\!=\!\!\mathbf{V}_s^{(\!-\!1\!)}\!(\!\tau\!)\!\circ\!{\mathbf{U}_{\! s}}\!(\!\tau\!).
\end{equation*}

\subsection{Complexity of BGMP}
The matrix form of BGMP permits a parallel processing and further reduces the complexity and latency. In each iteration, it costs about $20KM$ multiplications (or divisions) and $2KM$ exponents (or logarithms). If we use the approximated Bernoulli message update (\ref{ap_sum_ber}), the complexity can be further reduced to  $15KM$ multiplications and $KM$ exponents per iteration, which saves 25 percent of the multiplications and 50 percent of the exponents. Therefore, the complexity of BGMP is as low as $\mathcal{O}(KMN_{ite})$ multiplications and $\mathcal{O}(KMN_{ite})$ exponents, where $N_{ite}$ is the number of iterations. The scalar operation at each node in BGMP avoids the large-scale matrix calculations, which is the key reason resulting in a lower complexity of BGMP.

\subsection{GA-MMSE Bound of BGMP}
\textbf{\textit{Proposition 1:}} \emph{If the entries of $\mathbf{H}$ are \emph{i.i.d.} with a normalized distribution $\mathcal{N}(0,1/{K})$, the average \emph{MSE} of BGMP is bounded by that of the \emph{genie-aided MMSE} (GA-MMSE), i.e.,
\begin{equation}\label{bounds}
\mathrm{MSE}_{bgmp}>\mathrm{MSE}_{ga-mmse},
\end{equation}
where
\begin{equation} \mathrm{MSE}_{ga-mmse} \approx 1-0.25\mathcal{F}(\tfrac{\lambda K}{M},\sigma_n^{2}),
\end{equation}
and $\mathcal{F}(a,c)=\big(\sqrt{(1+1/\sqrt{a})^2 + c} -\sqrt{(1-1/\sqrt{a})^2+c}\,\,\big)^2$.}

\begin{IEEEproof}
Let $\mathbf{x}_+$ be the non-zero subvector of $\mathbf{x}$, and $\mathbf{H}_+$ the corresponding sub-measurement matrix of $\mathbf{H}$. Hence, $\mathbf{y}=\mathbf{H}_+\mathbf{x}_+ +\mathbf{n}$. Consider the following GA-MMSE method, where the non-zero index of $\mathbf{b}$ is known.
\begin{equation}
\hat{\mathbf{x}}_+=\big(\mathbf{H}_+^T\mathbf{H}_++\sigma_n^{2}\lambda \mathbf{I}\big)^{-1}\mathbf{H}_+^T\mathbf{y}.
\end{equation}
Obviously, the MSE of GA-MMSE is strictly less than that of BGMP, and thus is a strict lower bound of the MSE of BGMP. Similarly, the MSE of GA-MMSE is calculated by \cite{verdu2004}
\begin{equation}
\mathrm{MSE}_{ga\!-\!mmse}\!\!=\!\!\frac{1}{ K}\!\mathrm{Tr}\{\!\big(\sigma_n^{\!-2}\mathbf{H}_+^T\mathbf{H}_+ \!+\! \lambda \mathbf{I}\big)^{\!-\!1}\!\} \!\approx\!\! 1\!-0.25\mathcal{F}(\beta_+,\sigma_n^{2}),
\end{equation}
Hence, we have Proposition 1.
\end{IEEEproof}
\begin{figure}[t]
\centering
\includegraphics[width=8.2cm]{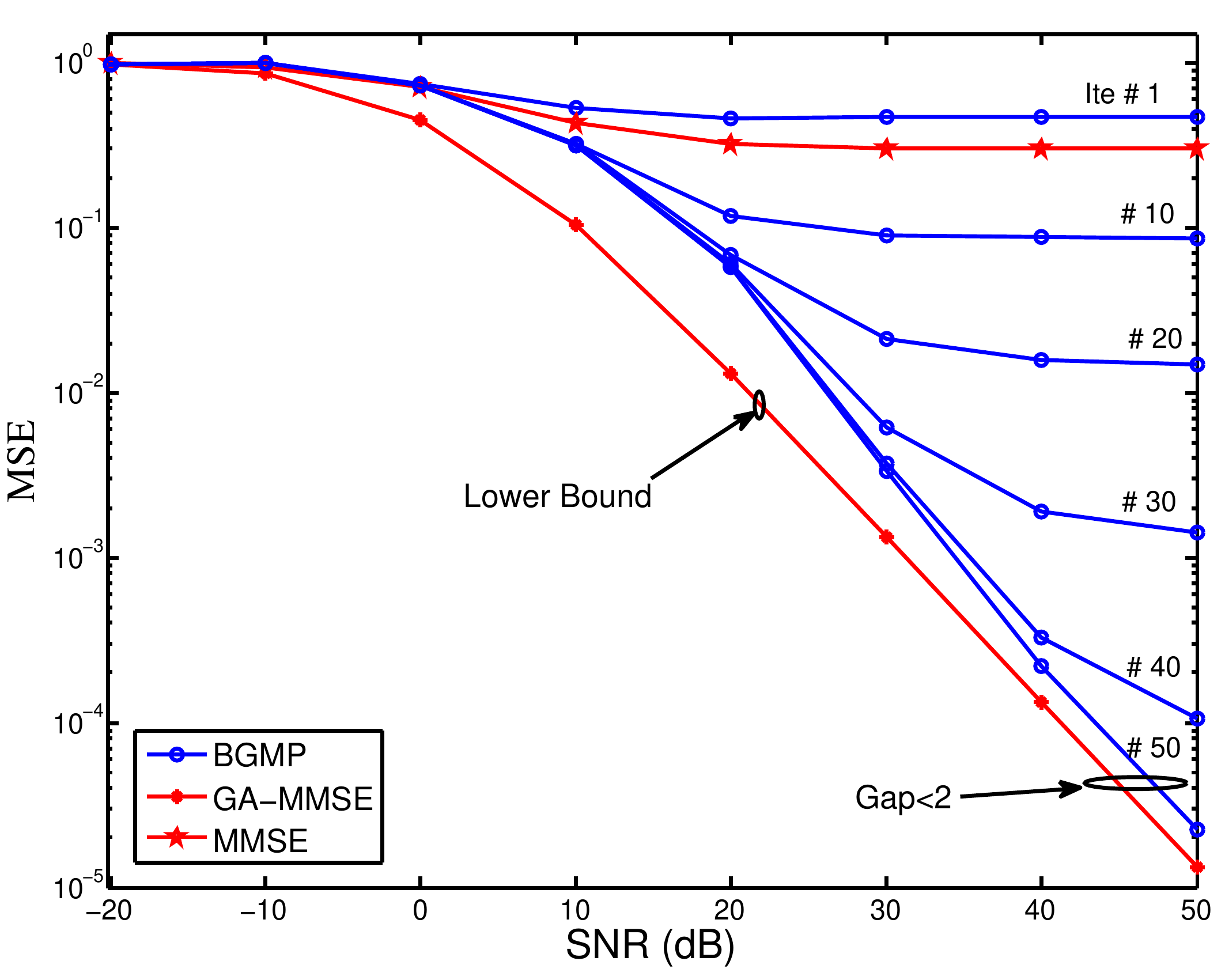}\\\vspace{-0.2cm}
    \caption{\footnotesize MSE comparison between the simulated BGMP, general MMSE and  genie-aided MMSE method (a lower bound assuming that the receiver knows the sparse position). $K=8192$, $M=5734(\approx0.7K), \mathrm{and}\;\lambda=0.4$. }\label{MSEfig}
\end{figure}
\begin{figure}[t]
\centering
\includegraphics[width=8.0cm]{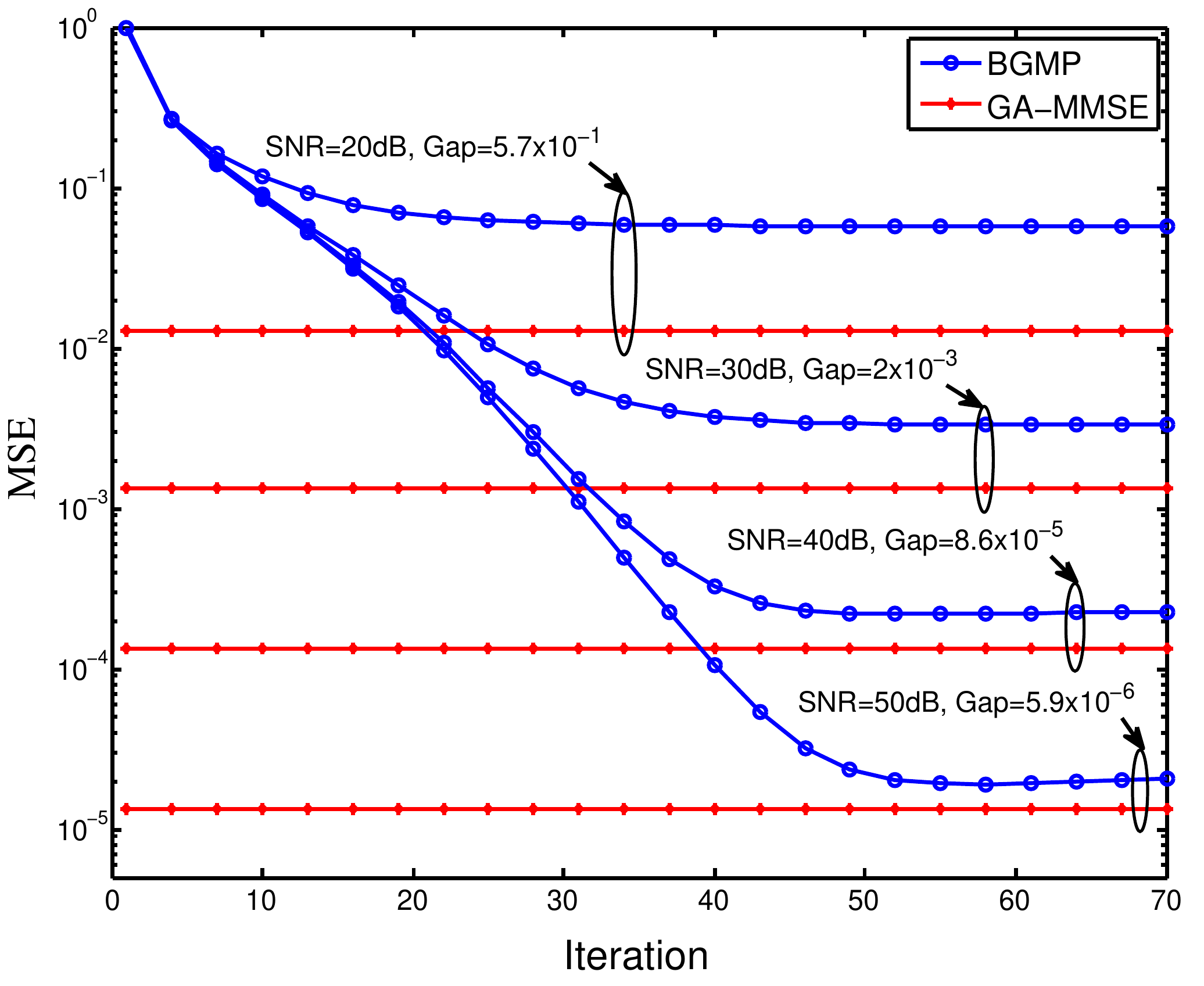}\\\vspace{-0.2cm}
    \caption{\footnotesize MSE comparison between the simulated BGMP and genie-aided MMSE method. $K=8192$, $M=5734(\approx\!\!0.7K), \mathrm{and}\; \lambda=0.4$.}
    \label{Ite_MSEfig}\vspace{-0.0cm}
\end{figure}
\section{Numerical Results}
In this section, we report the numerical results of the proposed BGMP for sparse vector recovery. For all experiments, we set signal-to-noise ratio to $\mathrm{SNR}=\frac{1}{\sigma_n^2}$, average SER to $\mathrm{SER}=\frac{1}{K}\cdot \|\mathbf{b}-\hat{\mathbf{b}}\|_1$, average MSE to $\mathrm{MSE}=\frac{1}{K}\cdot E[\|\mathbf{x}-\hat{\mathbf{x}}\|_2^2]$, and the entries of $\mathbf{H}$ are \emph{i.i.d.} with a normalized distribution $\mathcal{N}(0,1/{K})$. All the SERs and MSEs are averaged over 100 realizations.
\begin{figure}[t]
  \centering
  \includegraphics[width=8.5cm]{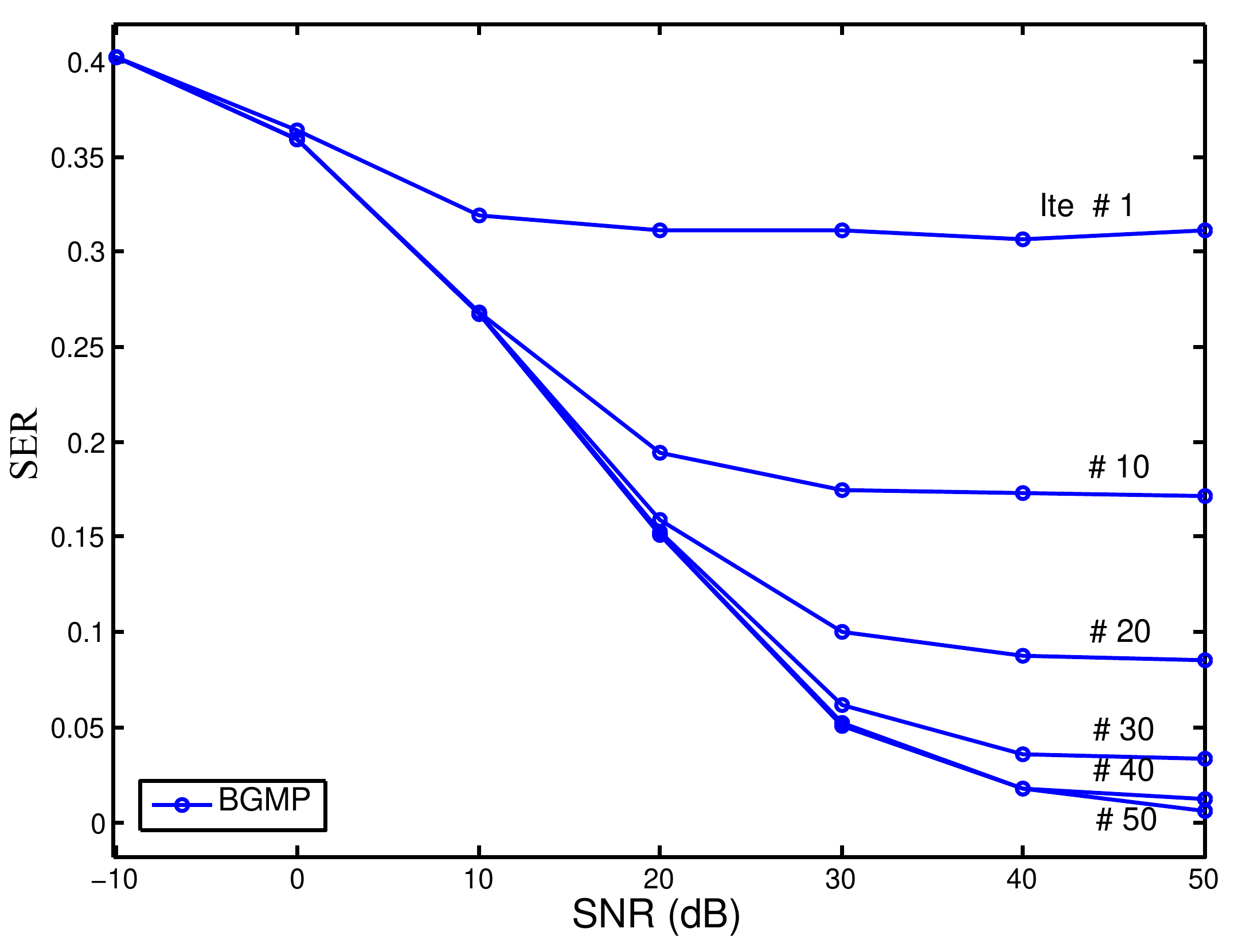}\\\vspace{-0.2cm}
        \caption{\footnotesize SERs of the GBMP. $K=8192$, $M=5734(\approx0.7K), \lambda=0.4, \mathrm{and}\; N_{ite}=1\sim50$.}\label{BERfig}
\end{figure}
\begin{figure}[t]\vspace{0.30cm}
  \centering
  \includegraphics[width=8.5cm]{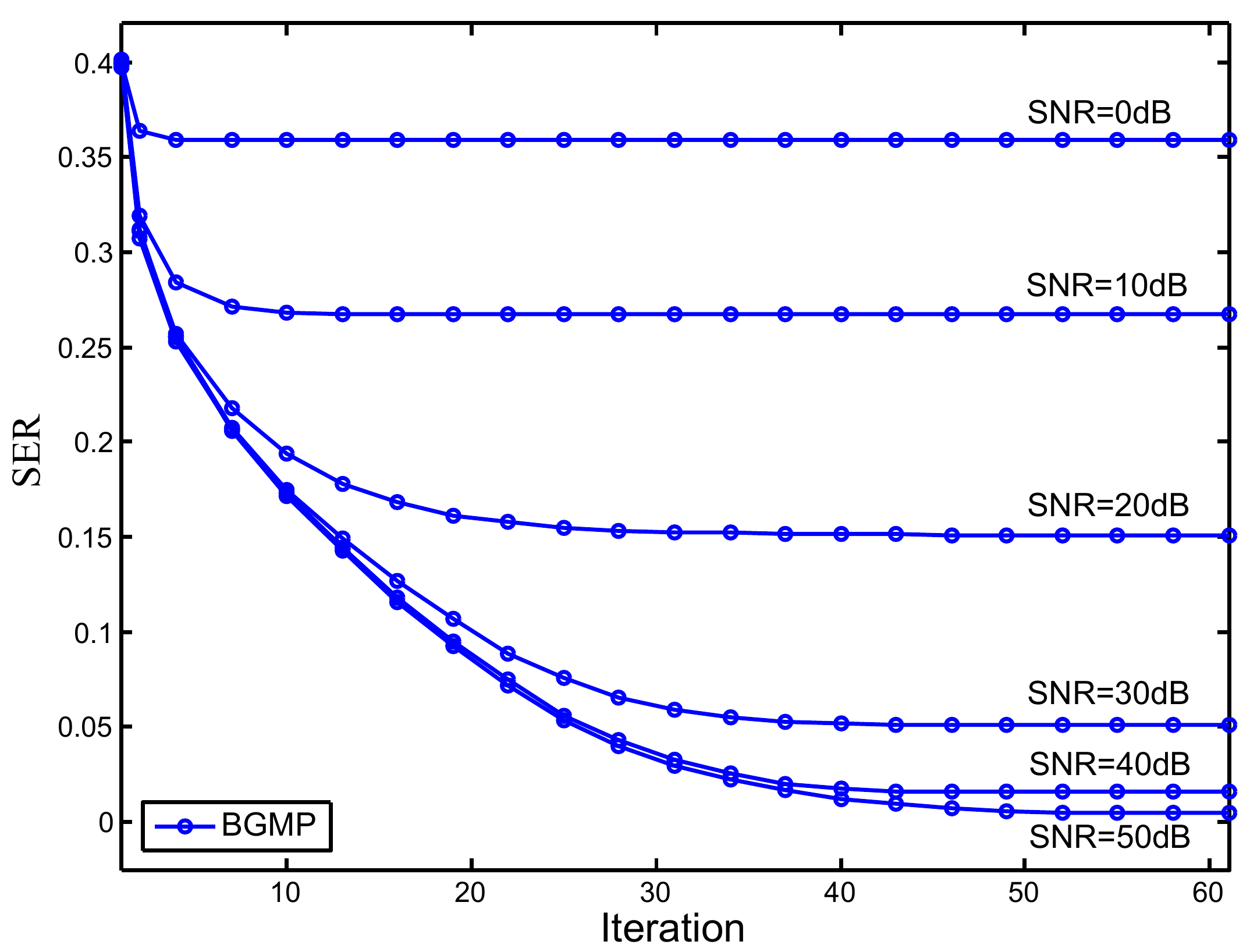}\\\vspace{-0.2cm}
        \caption{\footnotesize SERs of the GBMPA. $K=8192$, $M=5734(\approx0.7K), \lambda=0.4, \mathrm{and}\; N_{ite}=1\sim50$.}\label{Ite_BERfig}
\end{figure}\vspace{-0.2cm}
\begin{figure}[t]\vspace{-0.2cm}
  \centering
  \includegraphics[width=8.5cm]{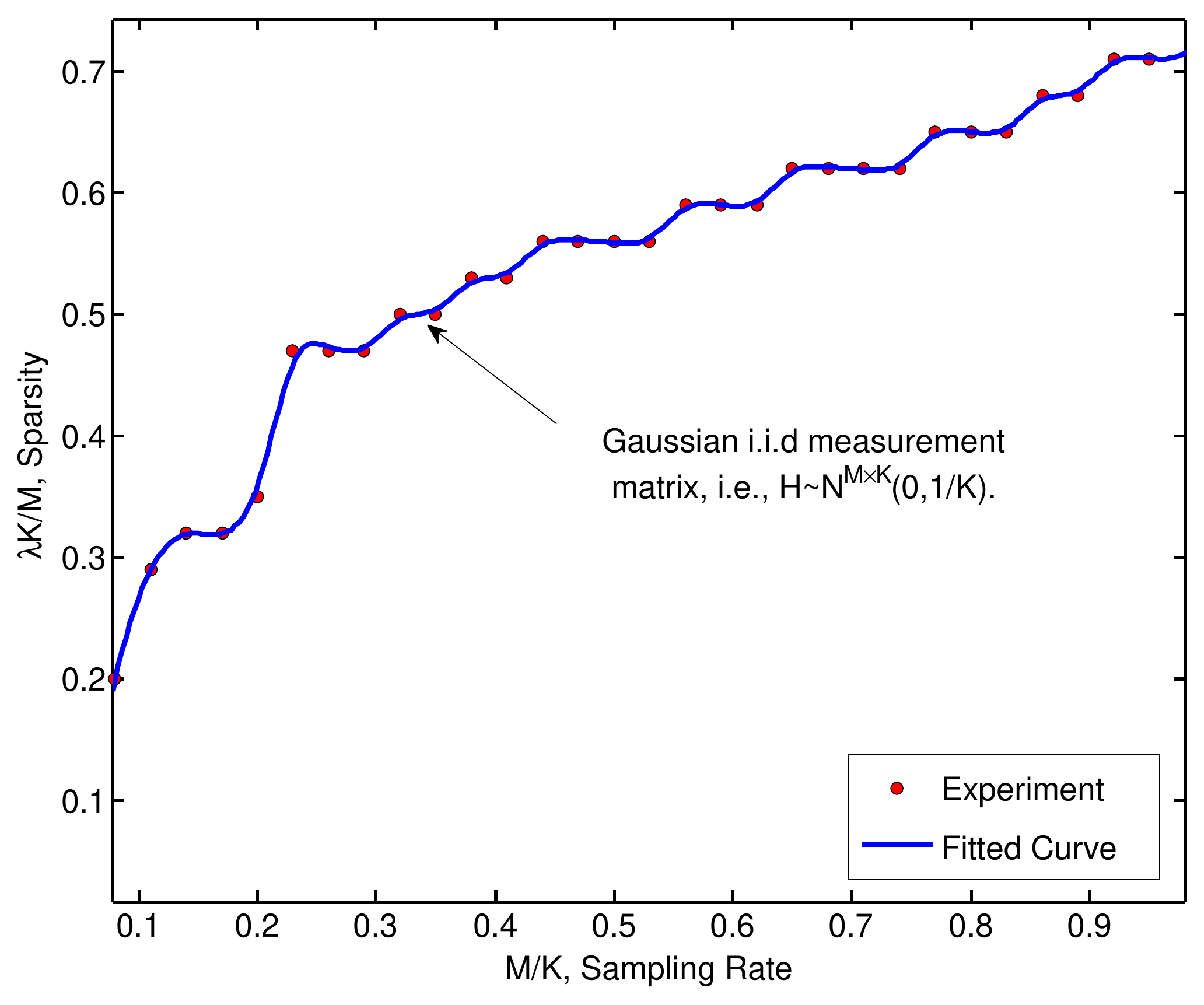}\\\vspace{-0.2cm}
        \caption{\footnotesize Noiseless empirical phase transition curve for BGMP algorithm with \emph{i.i.d.} Gaussian measurement matrix $\mathbf{H}\sim\mathcal{N}^{M\times K}(0,1/K)$, vector length $K=1000$, and $N_{ite}=100$ iterations. Here, $\lambda K$ is the number of nonzero components in $\mathbf{x}$. }\label{PTC}\vspace{-0.2cm}
\end{figure}
\subsection{MSE Performance of the Value Recovery}
In Fig. \ref{MSEfig}, we compare the MSE of the simulated BGMP, general MMSE and GA-MMSE, where $K=8192$, $M=5734(\approx0.7K)$, $\lambda=0.4$, and $N_{ite}=1\sim50$. We see that the proposed BGMP always outperforms the general MMSE method. In addition, after 50 iterations, the MSE of the proposed BGMP is approaching that of the GA-MMSE lower bound (the gap is less than 2dB) when $\mathrm{SNR}\geq40\mathrm{dB}$. Fig. \ref{Ite_MSEfig} presents the convergence of the BGMP under different $\mathrm{SNR}$s. It shows that the gap between MSE of BGMP and GA-MMSE decreases with the increase of SNR, and their gap is less than $6\times10^{-6}$ when $\mathrm{SNR}=50\mathrm{dB}$. Furthermore, the required the number iterations increases with SNR.

\subsection{SER Performance of the Support Recovery}
Fig. \ref{BERfig}-\ref{Ite_BERfig} show the SER of the simulated BGMP, where $K=8192$, $M=5734(\approx0.7K)$, $\lambda=0.4$, and $N_{ite}=1\sim50$. We see that after 50 iterations, the proposed BGMP recovers sparse positions with a very low error probability (less $10^{-2}$) when $\mathrm{SNR}\geq40\mathrm{dB}$. In addition, the SER decreases with the increase of SNR, and the required number of iterations increases with SNR.
\subsection{Noiseless Phase Transition}
The experiment results of the noiseless empirical \emph{phase transition curve} (PTC) are described in Fig. \ref{PTC}. To compute each empirical PTC, a $30\times 30$ grid of sampling rate $\frac{M}{K}\in[0.05,0.95]$ and sparsity $\frac{\lambda K}{M}\in[0.05,0.95]$ for fixed vector length $K=1000$ is constructed. At each grid point, we perform 100 independent realizations of a Bernoulli-Gaussian vector $\mathbf{x}$ and an $M\times K$ measurement matrix $\mathbf{H}$ with \emph{i.i.d.} $\mathcal{N}(0,K^{-1})$ entries. We consider the noiseless case that $\mathbf{y}=\mathbf{H}\mathbf{x}$, and the proposed BGMP is used for the recovery of vector $\mathbf{x}$. A recovery $\hat{\mathbf{x}}$ in each realization is defined a success if $MSE=\frac{1}{K} \|\hat{\mathbf{x}} -\hat{\mathbf{x}}\|_2^2<10^{-6}$, and the average success rate is defined as $P_s=\frac{S}{100}$, where $S$ is the number of success recovery in the 100 realizations. The empirical PTC is then plotted, using Matlab's contour command, as the contour $P_s=0.5$ over the sparsity-sampling grid.

\section{Conclusion}
In this paper, we have proposed a low-complexity BGMP algorithm for sparse vector recovery, where Gaussian messages and Bernoulli messages perform the value estimation and support estimation respectively. In addition, a GA-MMSE lower bound has been provided for the MSE of BGMP. Our numerical results showed the tightness of the GA-MMSE lower bound in high SNR, the excellent MSE performance in value recovery, and out-standing SER performance in support recovery. Particularly, the MSE curve of BGMP is less than 2dB away from the GA-MMSE lower bound at $\mathrm{MSE}=10^{-4}$, and less than $6\times10^{-6}$ away from the GA-MMSE lower bound at $\mathrm{SNR}=50\mathrm{dB}$. Besides, the SERs of the proposed BGMP is less than $10^{-2}$ when $\mathrm{SNR} \geq40 \mathrm{dB}$.

\end{document}